\documentclass[12pt]{article}
\usepackage{amsmath}
\usepackage{graphicx,psfrag,epsf}
\usepackage{enumerate}
\usepackage{algorithmic,algorithm}
\usepackage{natbib}
\usepackage{pbox}
\usepackage{makecell}
\usepackage{color}
\usepackage{url} 

\newcommand{\blind}{1}

\newtheorem{thm}{Theorem}

\newcommand{\bA}{\mathbf{A}}
\newcommand{\bB}{\mathbf{B}}
\newcommand{\bX}{X}
\newcommand{\bY}{Y}
\newcommand{\bw}{w}
\newcommand{\E}{\mathrm{E}}
\newcommand{\Var}{\mathrm{Var}}
\newcommand{\Cov}{\mathrm{Cov}}

\DeclareMathOperator*{\argmax}{\arg\!\max}

\DeclareMathOperator*{\logit}{\mathrm{logit}}

\begin{document}

	\def\spacingset#1{\renewcommand{\baselinestretch}%
		{#1}\small\normalsize} \spacingset{1}

	
	\if1\blind
	{
		\title{\bf A Tracy-Widom Empirical Estimator For Valid P-values With High-Dimensional Datasets}
		\author{Maxime Turgeon\thanks{
				Computations were made on the supercomputer Mammouth-parall\`ele 2 from Universit\'e de Sherbrooke, managed by Calcul Qu\'ebec and Compute Canada. The operation of this supercomputer is funded by the Canada Foundation for Innovation (CFI), the minist\`ere de l'\'Economie, de la science et de l'innovation du Qu\'ebec (MESI) and the Fonds de recherche du Qu\'ebec---Nature et technologies (FRQ-NT). 
				The methylation data used for the data analysis was kindly provided by Marie Hudson, Sasha Bernatsky, Ines Colmegna, and Tomi Pastinen.
				{Finally, the authors would also want to thank Stepan Grinek for bringing to our attention the machine learning and pattern recognition literature on the use of truncated SVD.}
			}\\
			Department of Epidemiology, Biostatistics and Occupational Health\\
			McGill University\\
			\\
			Celia Greenwood\\
			Department of Epidemiology, Biostatistics and Occupational Health\\
			McGill University\\
			and\\
			Aur\'elie Labbe\\
			Department of Decision Sciences, HEC Montr\'eal
}
		\maketitle
	} \fi
	
	\if0\blind
	{
		\bigskip
		\bigskip
		\bigskip
		\begin{center}
			{\LARGE\bf  A Tracy-Widom Empirical Estimator For Valid P-values With High-Dimensional Datasets}
		\end{center}
		\medskip
	} \fi
	
	\bigskip
	\if1\blind 
	{
		\clearpage
	} \fi
	
	\begin{abstract}
		Recent technological advances in many domains including both genomics and brain imaging have led to an abundance of high-dimensional and correlated data being routinely collected. Classical multivariate approaches like Multivariate Analysis of Variance (MANOVA) and Canonical Correlation Analysis (CCA) can be used to study relationships between such multivariate datasets. Yet, special care is required with high-dimensional data, as the test statistics may be ill-defined and classical inference procedures break down.
		
		In this work, we explain how valid p-values can be derived for these multivariate methods even in high dimensional datasets. Our main contribution is an empirical estimator for the largest root distribution of a singular double Wishart problem; this general framework underlies many common multivariate analysis approaches. From a small number of permutations of the data, we estimate the location and scale parameters of a parametric Tracy-Widom family that provides a good approximation of this distribution. Through simulations, we show that this estimated distribution also leads to valid p-values that can be used for high-dimensional inference. We then apply our approach to a pathway-based analysis of the association between DNA methylation and disease type in patients with systemic auto-immune rheumatic diseases.
	\end{abstract}
	
	\noindent%
	\vfill
	
	\newpage
	\spacingset{1.45} 
	
	\section{Introduction}
	
	Consider the following scenario: your research team received a small grant to study the relationship between anatomical brain features and a set of clinical phenotypes.  For the past several months, you've been painstakingly enrolling patients in the study and collecting neuroimaging data on them.  The subject-matter expert on your team has already grouped the millions of voxels into regions of interest.  A given region can now be represented by a dataset $Y$ of dimension $n\times p$, for $n$ patients, and the clinical phenotypes by a dataset $X$ of dimension $n\times q$.  Prior to the study, you had identified Canonical Correlation Analysis (CCA) as a potential analytical tool; it would allow the extraction of maximally correlated components from both datasets, and the overall relationship between $Y$ and $X$ could be summarised with a series of canonical correlations.  In the classical low-dimensional setting ($p < n, q< n$), you could also test for significance of these canonical correlations using Rao's statistic~\citep{mardia1979multivariate}.  However, despite your colleague's excellent work, most brain regions still contain more features or measurements than your overall sample size ($p > n$); in other words, you are no longer in the classical inference setting, but in a high-dimensional setting.  Building on your knowledge of linear algebra and matrix decomposition, you know that you can still extract the canonical components and canonical correlations using a truncated eigenvalue decomposition (EVD).  But one last obstacle remains: the null distribution of the first canonical correlation is unknown in this high-dimensional setting.  Although you could rely on a permutation strategy to obtain a p-value, you are also aware that such a procedure is very computationally  expensive if you want to reach a high level of precision for the estimated p-value, or if you want to analyze multiple regions in the brain.
	
	In this article, we provide a fast computational approach for estimating the null distribution of the first canonical correlation  in such  high dimensional setting.
	
	However, the contribution of this paper extends well beyond this CCA example provided above. More generally, a cursory look at the table of contents of recent volumes in both neuroimaging and genomics journals reveals a strong bias towards multivariate analysis methods such as Principal Component Analysis (PCA), Multivariate Analysis of Variance (MANOVA), Canonical Correlation Analysis (CCA), Principal Components of Explained Variance (PCEV), and Linear Discriminant Analysis (LDA); for a small yet broad sample, see \citet{park2017groupwise,zhao2017feature,hao2017identification,pesonen2017combined,gossmann2018fdr,fraiman2018anova,happ2018multivariate,yang2018exploration,turgeon2016principal}. This is hardly surprising, given the evolving nature of technological capabilities data and the complex underlying biological processes that are now measurable.  As described in our scenario above, using truncated matrix decomposition, we can often perform dimension reduction even in high dimensions.  But beyond dimension reduction, many classical multivariate approaches also aim at summarizing the relationship between two datasets $Y$ and $X$; in the list above, CCA, MANOVA and PCEV all have this common goal. Furthermore, this subset of methods  also provide a unified way of performing null hypothesis significance testing: they all rely on the largest root of a \emph{double Wishart problem}.  Specifically, the strength of the association between $Y$ and $X$ is measured in terms of the magnitude of the largest solution to the following determinantal equation:
	\begin{equation}\label{eqn:deteqn}
	\det\left(\bB - \lambda(\bA + \bB)\right) = 0.
	\end{equation} 
	where $A$ and $B$ are two independent random matrices, both following a Wishart distribution with the same scale matrix. The definition of $A$ and $B$ is formulated under the null hypothesis and is method-specific. Several specific examples will be given later in this paper. From the distribution of this largest root, we can then compute a p-value for the null hypothesis of interest. 
	
	More formally, in high-dimensional settings, when the sample size is smaller than the number of measurements, both matrices $\bA$ and $\bB$ can have singular distributions. This singularity leads to both computational challenges for estimation and theoretical challenges for inference. On the one hand, common estimators in the non-singular case can be ill-conditioned (or even undefined) for singular problems; on the other hand, classical asymptotic convergence results rely on large sample sizes and therefore may not directly apply to high-dimensional settings.
	
	In this work, we are interested in multivariate analysis involving two datasets, $\bY$ and $\bX$, such that the dimension of one or both matrices may be much larger than the sample size $n$. We posit that proper high-dimensional inference in several multivariate statistical methods such as CCA, MANOVA and PCEV, can be attained by studying the \emph{singular} double Wishart problem described above. Our main contribution is an empirical estimator of the distribution of the largest root that is applicable to the analysis of high-dimensional data. This estimate provides valid p-values by fitting a location-scale family of distributions to a small number of permutations of the original data. By a theorem of~\citet{johnstone2008multivariate}, this family of distributions is known to provide an excellent approximation in the non-singular case, and we provide empirical evidence that this good performance extends to the singular case.
	
	The rest of the article is structured as follows: in Section~\ref{s:examples}, we provide some detailed examples of double Wishart problems; these examples are later used to illustrate our approach. In Section~\ref{s:heuristic}, we describe our approach to approximating the distribution function of its largest root using an empirical estimator. Then, in Section~\ref{s:sim}, we investigate the numerical accuracy of the approximation and show how it leads to valid p-values. We further illustrate our approach through an analysis of the association between DNA methylation and disease type in patients with systemic auto-immune rheumatic diseases. Finally, in Section~\ref{s:ledoit}, we explain how our empirical estimator can be extended to accommodate linear shrinkage covariance estimators within the double Wishart setting. Our approach has already been implemented in two \texttt{R} packages: \texttt{pcev} and \texttt{covequal}; both packages are available on \texttt{CRAN}.
	
	\subsection*{Notation}
	
	{In what follows, $X$ and $Y$ will denote $n\times q$ and $n\times p$ matrices, respectively.} We also write $\bA\sim W_p(\Sigma, n)$ when $\bA$ is Wishart-distributed with parameters $p,n$ and scale matrix $\Sigma$. Recall that this is equivalent to having an $n\times p$ matrix $X$ where each row is independently drawn from a multivariate normal $N_p(0, \Sigma)$ and such that $\bA = X^T X$.
	
	\section{Examples of double Wishart problems}\label{s:examples}
	
	As stated above, the developments in our paper build on theory associated with double Wishart problems. Therefore, we start here by giving four examples of well known multivariate tests that are each double Wishart problems in order to emphasize the number of different applications of the theoretical results to follow. In each case, if the two matrices $\bA$ and $\bB$ are singular or ill-conditioned, then the theoretical results developed for double Wishart problems no longer apply. The four examples are one-way MANOVA, a test of equality of covariance matrices, CCA, and Principal Component of Explained Variance (PCEV). The first example is given for its simplicity and historical importance; the other three examples are used later to illustrate our approach. 
	
	\subsection{MANOVA}
	
	Suppose that we have a set of $n$ independent observations $\{Y_{ik}\}$, where $Y_{ik}\sim N_p (\mu_k, \Sigma)$ denotes the $i$-th observation of the $k$-th group, with $i = 1,\ldots,n_k$, $k=1,\ldots,K$. In MANOVA, we are interested in the null hypothesis of equality of means $H_0 : \mu_ 1 = \cdots = \mu_k$. First, for each group, we can form the sample mean $\bar{Y}_k$ and covariance matrix $S_k$. We then compute two basic quantities: 1) the within-group sum of squares $W = \sum_k n_k S_k$; and 2) the between-group sum of squares $B = \sum_k n_k (\bar{Y}_k - \bar{Y})(\bar{Y}_k - \bar{Y})^T$, where $\bar{Y}=n^{-1} \sum_{k=1}^K\sum_{i=1}^{n_k} Y_{ik}$ is the overall mean. Under $H_0$, the matrices $W$ and $B$ are independent and Wishart-distributed, with $W \sim W_p (\Sigma, n - K)$ and $B \sim W_p (\Sigma, k - 1)$. The union-intersection test of $H_0$ uses the largest root of $W^{-1} B$ as a test statistic or, equivalently, that of $(W + B)^{-1}B$~\citep[Chapter 12]{mardia1979multivariate}. In the notation of Equation~\ref{eqn:deteqn}, we therefore can express the union-intersection test as a double Wishart problem with $\bA = W$ and $\bB = B$. This test statistic is also known as Roy's largest root, and it is one of the standard tests in classical MANOVA.
	
	\subsection{Test of covariance equality}
	
	Suppose that independent samples $X, Y$ from two multivariate normal distributions $N_p(\mu_1 , \Sigma_1)$ and $N_p(\mu_2 , \Sigma_2)$ lead to covariance estimates $S_1,S_2$ which are independent and Wishart distributed on $n_1,n_2$ degrees of freedom, respectively. For example, we could take $S_i$ to be the maximum likelihood estimate of the covariance matrix based on $n_i + 1$ observations. We are interested in testing the null hypothesis $H_0 : \Sigma_1 = \Sigma_2$; this can done using the largest root of the determinantal equation~\ref{eqn:deteqn} as a test statistic for which we set $\bA = n_1 S_1$ and $\bB = n_2 S_2$~\citep[Chapter 10]{anderson2003introduction}.
	
	\subsection{CCA}
	
	Suppose we have two datasets $X,Y$ of dimension $n\times q$ and $n\times p$, respectively. Recall that CCA seeks to find linear combinations of $X$ and $Y$ that are maximally correlated with each other. Assume each row $(X_i,Y_i)$ of $(X,Y)$ has joint distribution $N_{q+p}(0, \Sigma)$, where $\Sigma$ has the form 
	$$\Sigma = \begin{pmatrix}
	\Sigma_X & \Sigma_{XY} \\
	\Sigma_{XY}^T & \Sigma_Y
	\end{pmatrix}.$$
	For simplicity, we first assume $\Sigma_X = \Sigma_Y$ (which implies that $q=p$). Under our normality assumption, a hypothesis test for independence between $X$ and $Y$ is equivalent to a hypothesis test for $\Sigma_{XY} = 0$.
	
	Write $P = Y (Y^T Y )^{-1} Y^T$ and let $P^\perp = I - P$. Using these quantities, we can define
	\begin{align*}
	\bA &= X^TP^\perp X,\\
	\bB &= X^TP X.
	\end{align*}
	Under our assumptions, we have $\bA \sim W_q(\Sigma_X, n - p)$ and $\bB \sim W_q(\Sigma_X, p)$. We can test the null hypothesis $H_0: \Sigma_{XY} = 0$ using as a test statistic the largest root of the double Wishart problem corresponding to the pair of matrices $\bA,\bB$ above~\citep[Chapter 10]{mardia1979multivariate}. We note that this largest root corresponds to the square of the first canonical correlation between $X$ and $Y$.
	
	The computation above can be generalized to the case when $\Sigma_X \neq \Sigma_Y$ (and $q\neq p$) and with nonzero mean; for details, see~\citet{johnstone2009approximate}.
	
	\subsection{PCEV}
	
	Similar to CCA, PCEV can also be used to simultaneously perform dimension reduction and test for association between two multivariate samples $X,Y$ of dimension $n\times q$ and $n\times p$, respectively. However, whereas CCA seeks to maximise the correlation between linear combinations of $Y$ and $X$, PCEV seeks the linear combination of $Y$ whose proportion of variance explained by $X$ is maximised. Specifically, we assume that the relationship between $X$ and $Y$ can be represented via a linear model:
	$$ Y = \Gamma X + E,$$
	where $\Gamma$ is a $p\times q$ matrix of regression coefficients for the covariates of interest, and $E\sim N_p(0, \Sigma_R)$ is a vector of residual errors. This model assumption allows us to decompose the total variance of $Y$ as the sum of variance explained by the model and residual variance:
	\begin{align*} 
	\Var(Y) &= \Sigma_M + \Sigma_R,
	\end{align*} 
	where $\Sigma_M = \Gamma\Var(X)\Gamma^T$. PCEV seeks a linear combination of outcomes, $\bw^TY$, that maximises the proportion $R^2(\bw)$ of variance being explained by the covariates $X$: 
	$$\bw_\mathrm{PCEV} = \argmax_{\bw \in \mathbf{R}^p : \|\bw\| = 1} R^2(\bw),$$ 
	where 
	\begin{equation}\label{eqn:h2}
	R^2(\bw) =\frac{ \Var(\bw^T\Gamma X)}{ \Var(\bw^TY)}=\frac{\bw^T\Sigma_M\bw}{\bw^T(\Sigma_M + \Sigma_R) \bw}.
	\end{equation}
	Then testing the null hypothesis $H_0 : \Gamma = 0$ is performed using as a test statistic the largest root of the determinantal equation~\ref{eqn:deteqn} for which we set $\bA = \Sigma_R$ and $\bB = \Sigma_M$. \citet{turgeon2016principal} have further details on this dimension-reduction technique.
	
	{For further examples of double Wishart problems, we refer the reader to~\citet{johnstone2008multivariate,johnstone2009approximate}.}
	
	\section{Empirical Estimator {of the Distribution of the Largest Root}}\label{s:heuristic}
	
	As we can see from the examples above, many null hypothesis significance tests in multivariate analysis follow the same two steps 1) computing the largest root of equation~\ref{eqn:deteqn}; and 2) from its distribution under the null hypothesis, compute a p-value. For the first point, we may have to use a truncated version of singular value decomposition (SVD) (or EVD) when both $\bA$ and $\bB$ are singular; this singularity is common in high-dimensional datasets. {In essence, these matrix decompositions are restricted to the subspace spanned by the eigenvectors corresponding to the non-zero eigenvalues; the mathematical details are reviewed in Section 1 of the Supplementary Materials.} In this section, we focus on the second point. We show how the distribution of the largest root can be accurately approximated in the singular setting.
	
	\subsection{Known results in the non-singular setting}
	
	In the non-singular setting, the distribution of the largest root to the determinantal equation~\ref{eqn:deteqn} is well studied~\citep{mardia1979multivariate,muirhead2009aspect}. To provide a theoretical underpinning to the remainder of the section, we highlight two approaches to computing this distribution: an exact approach, and an asymptotic result. 
	
	First, \citet{chiani2016distribution} described an explicit algorithm for computing the cumulative distribution function (CDF) of the largest root $\lambda$. Building on his earlier work~\citep{chiani2014distribution}, he proposed a new expression that relates the CDF to the Pfaffian of a skew-symmetric matrix. He also provided a set of recursive equations that provide a fast and efficient way to compute this matrix. However, this matrix quickly becomes very large when the parameters of the Wishart distribution increase, leading to both computational instability and numerical overflow problems. And yet, this matrix can only be computed in the non-singular setting.
	
	
	The second approach we wish to highlight uses results from random matrix theory to derive an approximation to the marginal distribution. Specifically, \citet{johnstone2008multivariate} showed that after a suitable transformation of the largest root, its distribution can be approximated by the Tracy-Widom distribution of order 1. His result, using the notation of this manuscript, is given below:
	
	\begin{thm}{[\citet{johnstone2008multivariate}]}\label{thm:johnApprox}
		Assume $\bA \sim W_p(\Sigma, m)$ and $\bB \sim W_p(\Sigma, n)$ are independent, with $\Sigma$ positive-definite.  Let $\lambda$ be the largest root of Equation~\ref{eqn:deteqn}. As $p,m,n\to\infty$, we have
		$$\frac{\logit\lambda - \mu}{\sigma} \stackrel{\mathcal{D}}{\longrightarrow} TW(1),$$
		where $TW(1)$ is the Tracy-Widom distribution of order 1 (cf.~\citet{tracy1996orthogonal}), and $\mu,\sigma$ are defined as follows: 
		\begin{align*}
		\mu &= 2 \log\left(\tan\left(\frac{\varphi + \gamma}{2}\right)\right)\\
		\sigma^3 &= \frac{16}{(m + n + 1)^2}\left(\sin^2(\varphi + \gamma)\sin\varphi\sin\gamma\right)^{-1},
		\end{align*} 
		where
		\begin{align*}
		\sin^2(\gamma/2) &= \frac{\min(p, n) - 1/2}{m + n + 1}\\
		\sin^2(\varphi/2) &= \frac{\max(p, n) - 1/2}{m + n + 1}.\\
		\end{align*}
	\end{thm}
	
	\subsection{Tracy-Widom Empirical Estimator}
	
	Unfortunately, in the singular setting, the marginal distribution of the largest root is not as well-understood. Crucially, the results from both~\citet{johnstone2008multivariate} and~\citet{chiani2016distribution} depend on the non-singularity of the matrix $\bA$, and therefore they cannot readily be applied to the singular setting.
	
	As highlighted in the introduction, a common approach to computing p-values when the null distribution is unknown is to use a permutation procedure. But high precision requires a large number of permutations, and therefore is computationally burdensome. In this section, we show that we can drastically reduce the number of permutations by using Johnstone's theorem above as inspiration. 

	We suggest an empirical estimator for approximating the CDF of the largest root. Indeed, we propose to estimate the distribution using a location-scale family of the Tracy-Widom distribution of order 1 indexed by two parameters $(\mu, \sigma)$. Algorithm~\ref{alg:empEstim} below describes our approach.
	
	\begin{algorithm}
	\caption{Tracy-Widom empirical estimator of the largest root distribution.}
	\label{alg:empEstim}
	\begin{algorithmic}[1]
		\STATE For a double Wishart problem associated with matrices $\bY$ and $\bX$:
		\FOR{$k=1$ to $K$}
		\STATE Perform a permutation procedure on $\bY$ and $\bX$.
		\STATE Compute the largest root $\lambda^{(k)}$ of the corresponding double Wishart problem.
		\ENDFOR
		\STATE Transform the roots $\lambda^{(1)}, \ldots, \lambda^{(K)}$ to the logit scale.
		\STATE Using a fitting procedure, find the estimates $\hat{\mu}$ and $\hat{\sigma}$ of the location and scale parameters, respectively.
		\STATE Approximate the CDF of the largest root using the CDF of $\hat{\sigma} TW(1) + \hat{\mu}$.
	\end{algorithmic}	
	\end{algorithm}
	
	A few comments are required:
	\begin{itemize}
		\item As we will show in Section~\ref{s:sim}, the number of permutations $K$ can be chosen to be relatively small while retaining good performance.
		\item The appropriate permutation procedure on Line 3 depends on the particular double Wishart problem being studied. For a test of association between $\bY$ and $\bX$, we typically permute the rows of $\bY$ and keep those of $\bX$ fixed. The test of equality of covariance matrices requires a different strategy, and we give all the relevant details in Section~\ref{s:covequal} below.
		\item Line 7 of Algorithm~\ref{alg:empEstim} refers to a fitting procedure. In Section~\ref{ss:approx}, we investigate four different approaches: Method of moments; Maximum Likelihood Estimation; and Maximum Goodness-of-Fit estimation~\citep{luceno2006fitting} using the Anderson-Darling statistic and a modified version that gives more weight to the right tail. Details about this latter approach, including computational formulas of these statistics, are given in Section 2 of the Supplementary Materials.
	\end{itemize}
	
	Therefore, we propose this location-scale transformation of the Tracy-Widom distribution as a suitable parametric family for estimating the distribution of the largest root in singular double Wishart problems.
	
	\section{Simulation results}\label{s:sim}
	
	To investigate the performance of our empirical estimator, we performed two simulation studies: 1) we compared the distribution obtained from our empirical estimator to an empirically generated cumulative distribution function (CDF), as described below, for different number of permutations $K$ and using four different fitting strategies; 2) we compared p-values derived from the estimated distribution to those obtained through a permutation procedure. The first simulation study is aimed at assessing the performance of our empirical estimator over the whole range of the distribution; the second simulation study specifically looks at the upper tail of the distribution and therefore at the validity of the resulting p-values. 
	
	\subsection{Comparison to the true distribution}\label{ss:approx}
	
	{Since the distribution function of the largest root distribution is not available analytically, we cannot use an analytical function as the benchmark for the ``true'' CDF. Therefore, an estimate of the true distribution was derived computationally.} Specifically, we started by generating 1000 pairs of singular Wishart variates $\bA \sim W_p(\Sigma, m),\bB \sim W_p(\Sigma, n)$ as follows: each singular Wishart variate was generated by first generating a sample of multivariate normal variates $Z_1, \ldots, Z_m \sim N_p(0, \Sigma)$, and then defining $\bA = \sum Z_i^TZ_i$; the Wishart variate $\bB$ was generated similarly. For each pair of Wishart variates, we then computed its corresponding largest root using truncated SVD (cf.\ \S 1 of the Supplementary Materials). From these 1000 largest roots, we calculated the empirical CDF for the marginal distribution: this estimate was considered our benchmark for assessing the performance of our approach.
	
	For our simulations, we fixed the degrees of freedom $m=96$ and $n=4$. In the context of a one-way MANOVA, this would correspond to four distinct populations and a total sample size of 100; in the context of CCA, this would correspond to one set of variates of dimension 4 (the dimension of the other is $p$), and a total sample size of 99. We also fixed the scale matrix $\Sigma = I_p$, but we varied the parameter $p=500,2000$. 
	
	To compute the Tracy-Widom empirical estimator, we sampled $K$ of these largest roots (with $K=25$ or $100$) and fitted the location-scale family as described in Algorithm~\ref{alg:empEstim}. Finally, we looked at four different fitting strategies: the method of moments (MM); Maximum Likelihood Estimation (MLE); and Maximum Goodness-of-Fit estimation~\citep{luceno2006fitting} using the Anderson-Darling statistic (AD) and a modified version that gives more weight to the right tail (ADR).
	
\begin{figure}
		\centering
		\includegraphics[width=\linewidth]{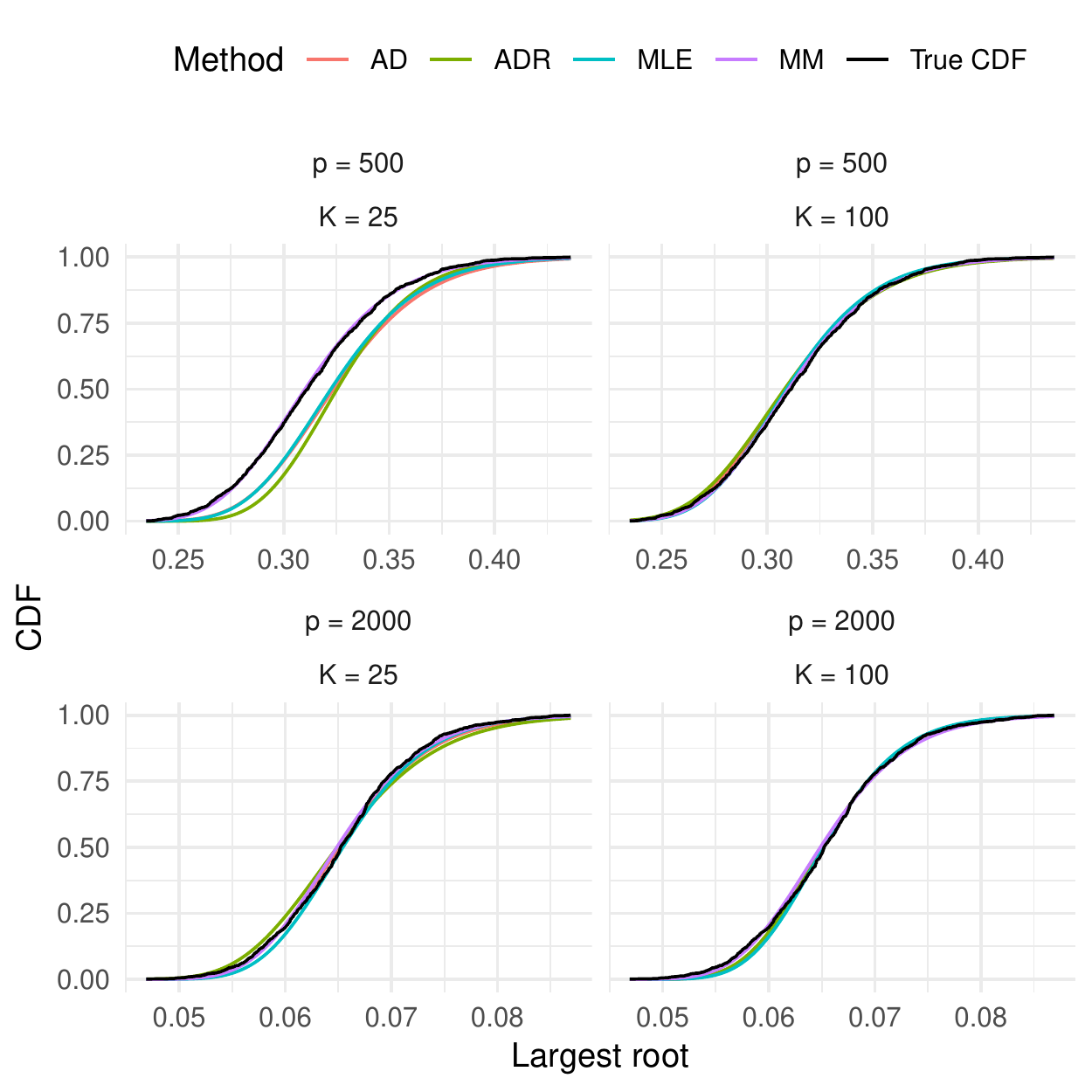}
		\caption{\textbf{Approximation to the CDF:} Tracy-Widom empirical estimator, using four different fitting strategies, compared to the ``true CDF'' (derived through computational means). AD: Anderson-Darling; ADR: right-tail-weighted Anderson-Darling; MLE: Maximum Likelihood Estimation; MM: Method of Moments.}
		\label{fig:approx}
	\end{figure}

	The simulation results appear in Figure~\ref{fig:approx}. As we can see, with a larger number of samples $K$, all four methods estimate the distribution of the largest root reasonably well; on the other hand, for a smaller value of $K$, the method of moments clearly outperforms the other fitting strategies. For this reason, unless otherwise stated, we use the method of moments in the remainder of this article.
	
	\subsection{Comparison of p-values}
	
	Next, we used our empirical estimator to compute p-values for three double Wishart problems: a test of equality of covariances, CCA, and PCEV. In all settings, we performed 100 simulations, and we fixed the sample size at $n=100$. We also used $K=100$ permutations to fit the Tracy-Widom empirical estimator using the method of moments. Finally, we compared our approach to a permutation procedure with 500 permutations. As a reference, we summarise the parameters for all simulation scenarios appear in Table~\ref{tab:sim}.
	
	\begin{table}[ht]
		\centering
		\begin{tabular}{l|cccc}
		\thead{Method} & \thead{Dimension $q$ of $X$} & \thead{Dimension $p$ of $Y$} & \thead{Association\\structure} & \thead{Association\\parameter} \\ \hline
		\makecell{Equality of\\covariance} & $200,300,400,500$ & $200,300,400,500$ & {\footnotesize \makecell{$\Sigma_X = I_q$\\Autoregressive $\Sigma_Y$}} & $\rho = 0,0.2,0.5$\\
		CCA & $200,300,400,500$ & Fixed at 20 & {\footnotesize \makecell{$(\Sigma_{XY})_{ii}=\rho$ for $i=1,2$,\\$(\Sigma_{XY})_{ij}=0$ otherwise}} & $\rho = 0,0.2,0.5$\\
		PCEV & Fixed at 1 & $200,300,400,500$ & Linear model & $R^2 = 0\%,1\%,5\%$ \\
		\end{tabular}
		\caption{Values of the parameters for all simulation scenarios} 
		\label{tab:sim}
	\end{table}
	
	\subsubsection{Test of equality of covariances}\label{s:covequal}
	For the test of equality of covariances, we simulated two datasets $X\sim N_p(0, I)$ and $Y\sim N_p(0, \Sigma)$, both with $n=100$ observations. We selected an autocorrelation structure for the covariance matrix $\Sigma$, with $\Cov(Y_i, Y_j) = \rho^{\lvert i - j\rvert}$. We varied two parameters: 1) the dimension $p=200, 300, 400, 500$ of both $X$ and $Y$; and 2) the correlation parameter $\rho=0, 0.2, 0.5$. Note that $\rho = 0$ corresponds to the null hypothesis of the same covariance, and $\rho=0.2,0.5$ correspond to alternative hypotheses.
	
\begin{figure}
		\centering
		\includegraphics[width=\linewidth]{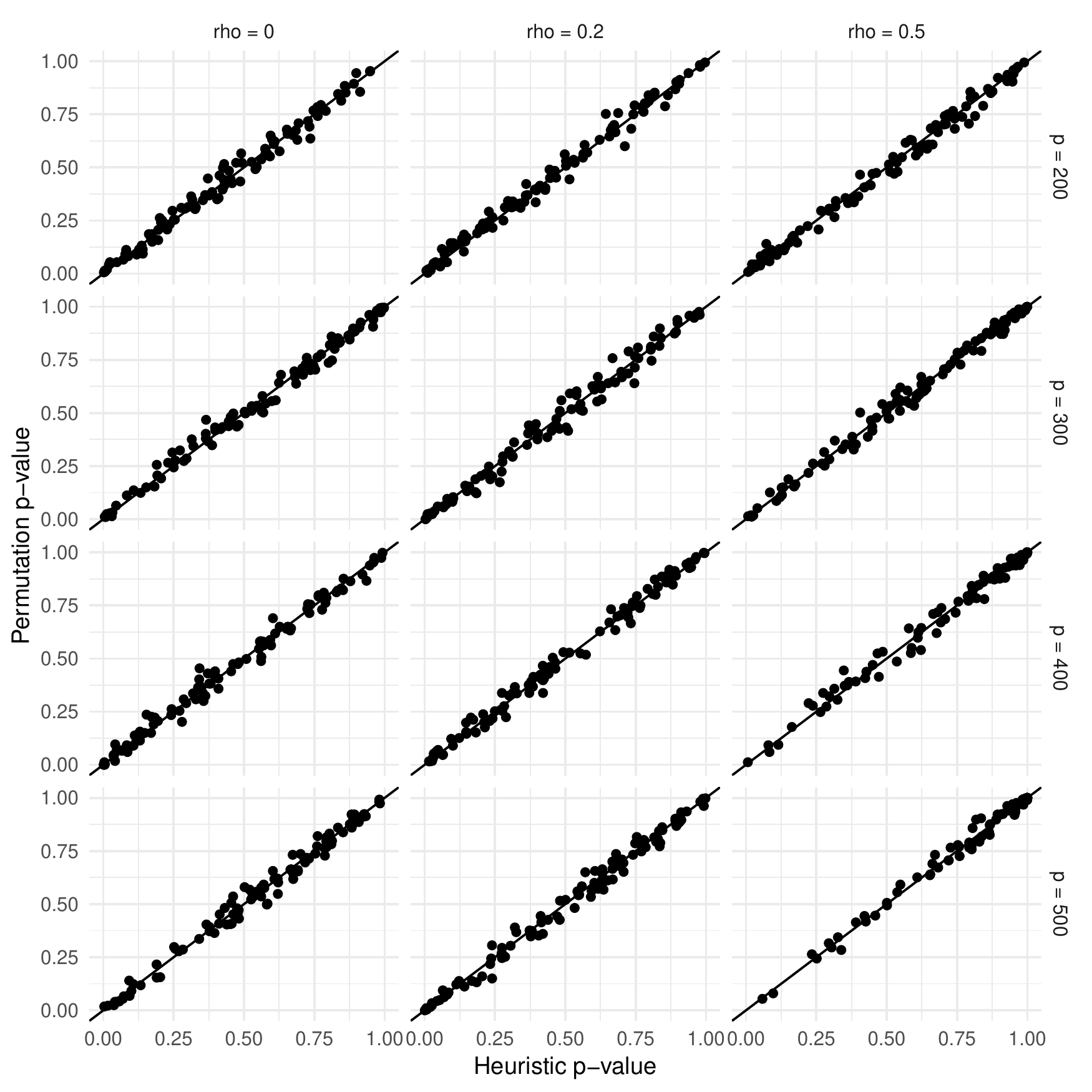}
		\caption{\textbf{Equality of covariance:} QQ-plots comparing the p-values obtained from a permutation procedure to those obtained from the Tracy-Widom empirical estimator.}
		\label{fig:var}
	\end{figure}

	The permutation procedure for testing the equality of covariance matrices started by centring both $X$ and $Y$. Then, the observations were permuted \emph{between} $X$ and $Y$: that is, a valid permutation would sample 50 observations from both $X$ and $Y$ to create a permuted $X$, and the remaining 50 observations from $X$ and $Y$ were used to create a permuted $Y$. The results are summarised using QQ-plots (see Figure~\ref{fig:var}). The computations were performed using the \texttt{R} package \texttt{covequal}.
	
	The simulation results appear in Figure~\ref{fig:var}. As we can for these QQ-plots, there is excellent agreement between the p-values obtained from a permutation procedure and those obtained from the Tracy-Widom empirical estimator.
	
	\subsubsection{CCA}
	For CCA, we again simulated two datasets $X\sim N_q(0, I)$ and $Y\sim N_p(0, I)$, both with $n=100$ observations, and with fixed $q = 20$. We selected an exchangeable structure with parameter $\rho$ for the covariance matrix $\Cov(X, Y)$. We again varied two parameters: 1) the dimension $p=200, 300, 400, 500$ of $X$; and 2) the correlation parameter $\rho=0, 0.2, 0.5$. As above, the value $\rho=0$ corresponds to the null hypothesis of no association between $X$ and $Y$, and the values $\rho=0.2,0.5$ correspond to alternative hypotheses. Finally, the permutation procedure consisted in permuting the rows of $X$ and keeping those of $Y$ fixed. 
	
\begin{figure}
		\centering
		\includegraphics[width=\linewidth]{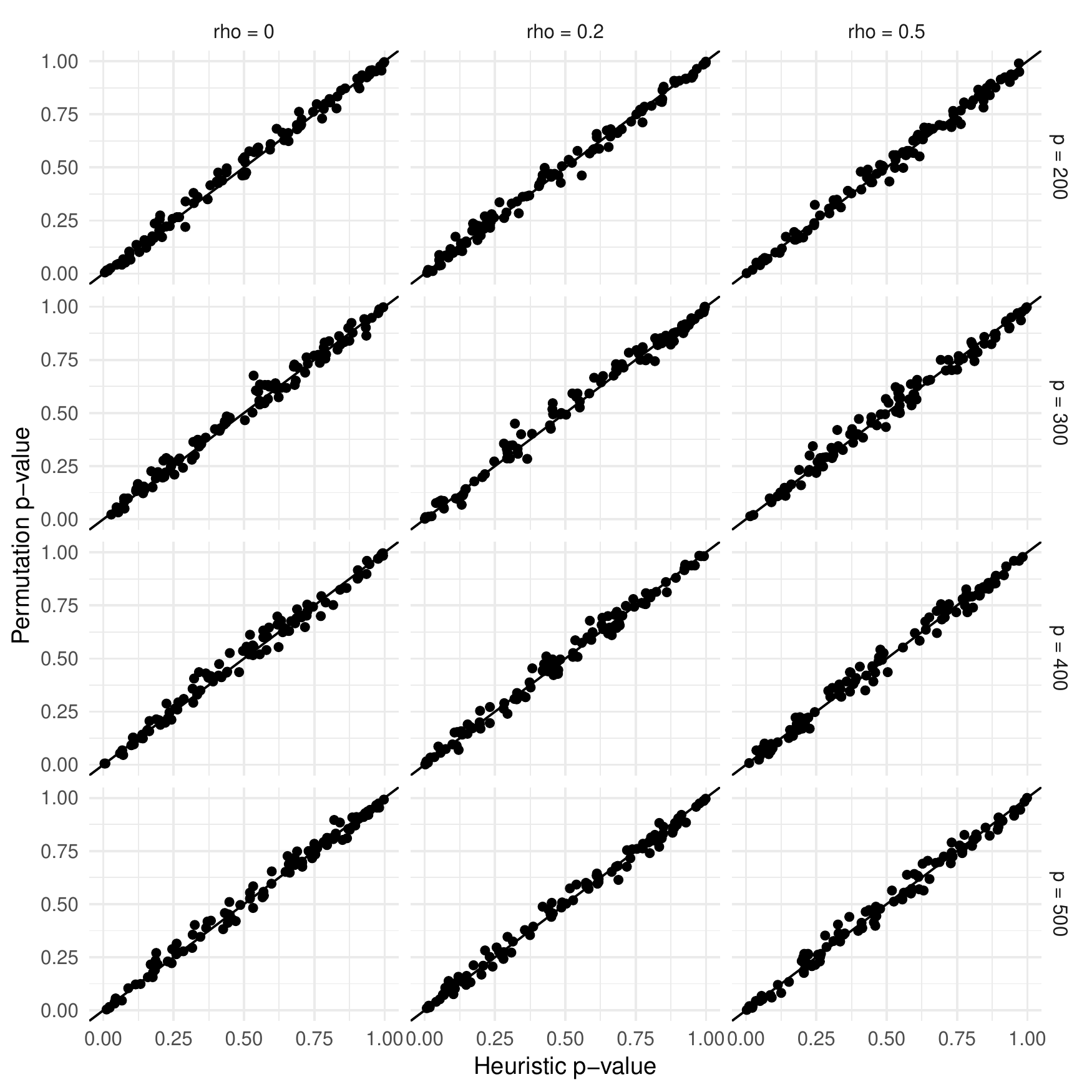}
		\caption{\textbf{CCA:} QQ-plots comparing the p-values obtained from a permutation procedure to those obtained from the Tracy-Widom empirical estimator.}
		\label{fig:cca}
	\end{figure}

	The simulation results appear in Figure~\ref{fig:cca}. As above, there is excellent agreement between the p-values obtained from a permutation procedure and those obtained from the Tracy-Widom empirical estimator.
	
	\subsubsection{PCEV}
\begin{figure}
		\centering
		\includegraphics[width=\linewidth]{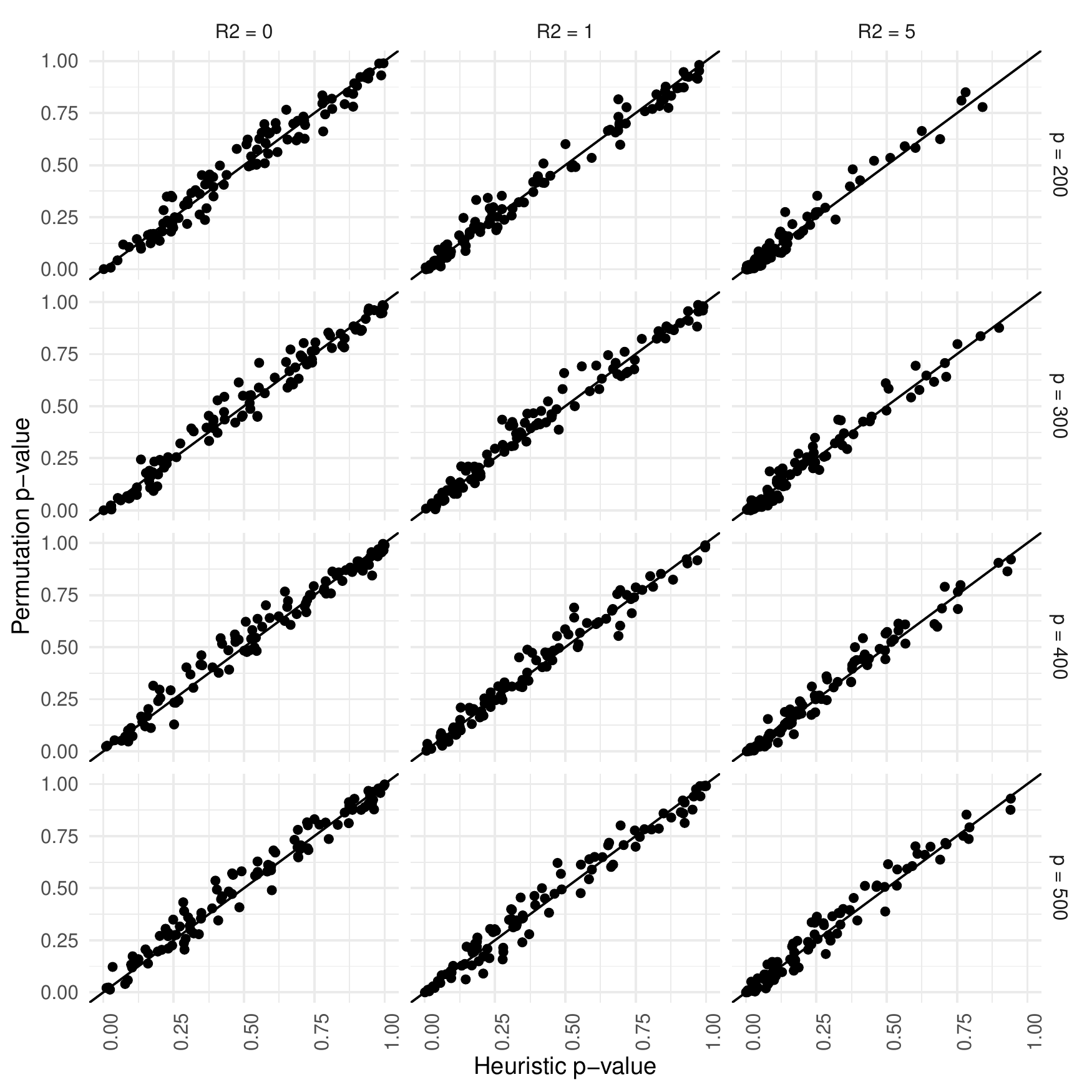}
		\caption{\textbf{PCEV:} QQ-plots comparing the p-values obtained from a permutation procedure to those obtained from the Tracy-Widom empirical estimator.}
		\label{fig:pcev1}
	\end{figure}

	For PCEV, we looked at the following high-dimensional simulation scenario: we fixed the number of observations $n=100$ and a balanced binary covariate $X$. We then varied the number of response variables $p=200, 300, 400, 500$, and fixed the covariance structure of the error term $\Sigma_R = I_p$. We induced an association between $X$ and the first 50 response variables in $Y$. This association was controlled by the parameter $R^2=0\%,1\%,5\%$; this parameter is related to the univariate regression coefficient $\beta$ through the following relationship:
	$$\beta^2 = \frac{R^2}{1-R^2}.$$
	As above, we summarised the results using QQ-plots (Figure~\ref{fig:pcev1}). The computations were performed using the \texttt{R} package \texttt{pcev}~\citep{turgeon2016principal}; the methodology presented here is part of the package starting with version 2.1.
	
	The simulation results appear in Figure~\ref{fig:pcev1}. Once again, there is excellent agreement between the p-values obtained from a permutation procedure and those obtained from the Tracy-Widom empirical estimator.
	
	As we can see, our Tracy-Widom empirical estimator yields valid p-values in a variety of high-dimensional scenarios that include both null and alternative hypotheses. For further simulation results, see Section 3 of the Supplementary Materials.
	
	\section{Data analysis}
	
	To showcase our ideas in the context of a real analysis of a high-dimensional dataset, we decided to look at the association between DNA methylation and disease type in patients with four systemic auto-immune rheumatic diseases: Scleroderma, Rheumatoid arthritis, Systemic lupus erythematosus, and Myositis. DNA methylation is an epigenetic mark, meaning that it is a chemical modification of the DNA that does not alter the nucleotide sequence~\citep{baylin2005dna}. It is known to be associated with changes in RNA transcription, and it is therefore correlated with gene expression. 
	
	The DNA methylation used for this analysis was measured prior to treatment on T-cell samples from 28 patients using the Illumina 450k platform~\citep{hudson2017novel}. Baseline characteristics of the patients appear in Table~\ref{tab:base}.
	
	\begin{table}[ht]
		\centering
		\begin{tabular}{l|cc}
			& Scleroderma & Other diseases \\
			& (n = 14) & (n = 14)\\ \hline
			Age {\tiny Mean (sd)} & 48 (16) & 52 (14) \\
			Female (\%) & 50\% & 71\% \\
		\end{tabular}
		\caption{Baseline characteristics} 
		\label{tab:base}
	\end{table}
	
	We opted to test for differential methylation between scleroderma and the other three diseases \emph{at the pathway level}: from the Kyoto Encyclopedia of Genes and Genomes (KEGG) \citep{kanehisa2017kegg}, we extracted the list of genes included in their manually curated list of molecular pathways; these pathways correspond to networks of genes interacting and reacting together as part of a given biological process. For each of the 320 pathways, we then identified all CpG dinucleotides contained in at least one gene of this pathway. All CpG dinucleotides mapped to a given pathway were analysed jointly. The extraction of gene lists was performed using the \texttt{R} package \texttt{KEGGREST}~\citep{keggrest}. 
	
	For each pathway, we thus have two datasets: an $n\times p$ matrix $Y$ that contains the methylation values at all $p$ CpG dinucleotides (where $p$ ranges from 39 to 21,640 over the 320 pathways) and an $n \times 1$ matrix $X$ indicating whether an individual has scleroderma. Recall that $n=28$, and therefore all pathways lead to a high-dimensional dataset. The analysis performed had two steps: first, we did a test of equality for the covariance matrices between both disease groups; then we used PCEV to test for differential methylation between these two groups. The PCEV analysis included both age and sex as possible confounders~\citep{el2007gender,horvath2013dna}. For both steps, the Tracy-Widom empirical estimator was computed using 50 permutations of the data.
	
	Since we repeated the same analysis independently for all 320 pathways, we need to account for multiple comparison. However, since a given gene may appear in multiple pathways, the 320 hypothesis tests are not independent; therefore, a naive Bonferroni correction would be too conservative. To estimate the effective number of independent tests, we looked at the average number of pathways in which a given CpG dinucleotide appears. Overall, 134,941 CpG dinucleotides were successfully matched to at least one of 320 KEGG pathways. On average, each dinucleotide appeared in 4.5 pathways; this leads to effectively 70 independent hypothesis tests. For a nominal family-wise error rate of $\alpha = 0.05$, an appropriate Bonferroni-corrected significance threshold is therefore given by $7.14 \times 10^4$.
	
	We compared the p-value obtained from our procedure above to that obtained from a permutation procedure; the latter is a common approach when the null distribution of a test statistic is unknown but has the disadvantage of being computationally expensive. Given our significance threshold, we determined that we needed to perform at least 10,000 permutations in order to be able to assess significance. 
	
	In Table~\ref{tab:results}, we present the five most significant pathways, with the top pathway achieving overall significance for the test of differential methylation. None of the covariance equality tests yielded a significant p-value; to improve clarity, we omitted these results from the table.

	\begin{table}[ht]
		\centering
		\begin{tabular}{llccc}
			\hline
			KEGG code & Description & \# CpGs & \pbox{3cm}{Tracy-Widom\\P-value} & \pbox{3cm}{Permutation\\P-value}\\ 
			\hline
			path:hsa00120 & Glutamatergic synapse & 225 & $1.91 \times 10^{-4}$ & $7.00 \times 10^{-4}$ \\ 
			path:hsa03040 & Ras signaling pathway & 2119 & $1.33 \times 10^{-3}$ & $1.40 \times 10^{-3}$ \\ 
			path:hsa03450 & Circadian rhythm & 267 & $1.52 \times 10^{-3}$ & $1.00 \times 10^{-4}$ \\ 
			path:hsa00563 & Histidine metabolism & 394 & $1.59 \times 10^{-3}$ & $3.00 \times 10^{-4}$ \\ 
			path:hsa04380 & Pathogenic E. coli infection & 2312 & $1.65 \times 10^{-3}$ & $5.20 \times 10^{-3}$ \\ 
			\hline
		\end{tabular}
		\caption{\textbf{Data analysis}: Five most significant pathways based on our empirical estimator. The Tracy-Widom p-values are compared to p-values obtained from a permutation procedure.} 
		\label{tab:results}
	\end{table}
	
\begin{figure}
		\centering
		\includegraphics[width=\linewidth]{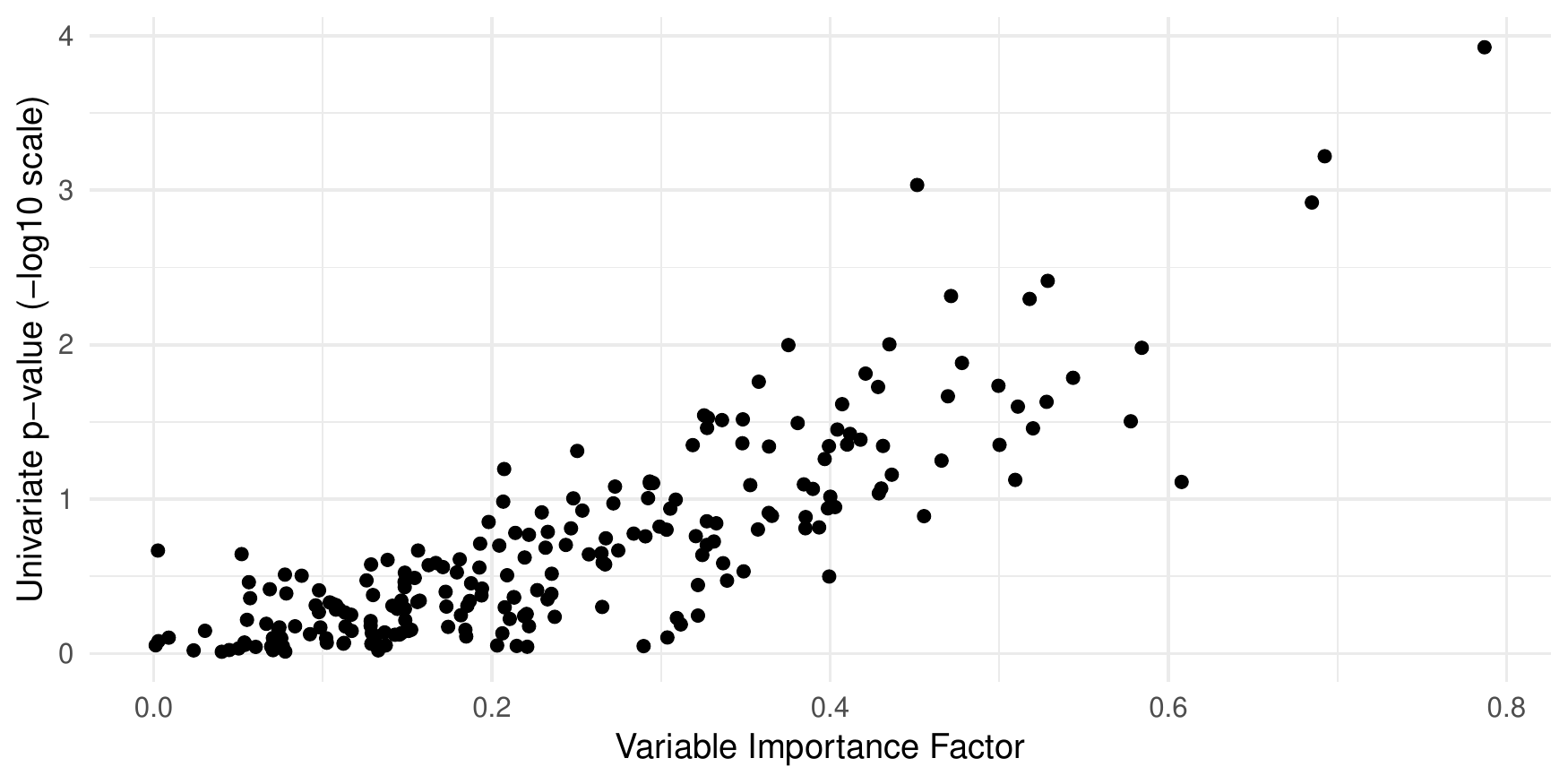}
		\caption{\textbf{path:hsa00120---Glutamatergic synapse}: Comparison of Variable Importance Factor and 225 univariate p-values for the most significant pathway.}
		\label{fig:vifVsUni}
	\end{figure}

	For the most significant pathway (i.e.\ Glutamatergic synapse), we computed the Variable Importance Factors (VIF) and compared them to the p-values obtained from a univariate approach where each CpG dinucleotide is tested individually against the disease outcome~\citep{turgeon2016principal}. The VIF is defined as the correlation between the individual response variables and the principal component maximising the proportion of variance. The comparison is presented in Figure~\ref{fig:vifVsUni}. As previously showed in the literature~\citep{turgeon2016principal}, there is some degree of agreement between these two measures of association. Moreover, we can see evidence that the overall association between this pathway and our disease indicator is driven by a few CpG dinucleotides.
	
	\section{Extension to linear shrinkage covariance estimators}\label{s:ledoit}
	
	In Section~\ref{s:sim}, we presented graphical evidence that our proposed Tracy-Widom empirical estimator provides a good approximation of the distribution of the largest root of the determinantal equation~\ref{eqn:deteqn}. As discussed in Section~\ref{s:examples}, the estimates of the matrices $\bA,\bB$ appearing in this equation often involve high-dimensional covariance matrices. However, a common problem with such high-dimensional covariance matrices is their instability~\citep[Chapter 1]{pourahmadi2013high}. {As a result, the power of statistical tests derived from the double Wishart problem decreases as the dimension of $\bA$ and $\bB$ increases. However, it would seem that the Tracy-Widom empirical estimator relies on the assumption that $\bA$ and $\bB$ are Wishart-distributed, and it is not clear \emph{a priori} that this estimator can be applied with other, more efficient estimators of the underlying high-dimensional covariances matrices.}
	
	One strategy for improving the stability of a covariance estimator is to use a shrinkage estimator. One such estimator for the covariance matrix was introduced by~\citet{ledoit2004well}. In this section, we show that by replacing the matrix $\bA$ by a linearly shrunk version $\bA^*$ in Equation~\ref{eqn:deteqn}, our Tracy-Widom empirical estimator still provides a good approximation of the distribution of the largest root. 
	
	Let $\bA\sim W_p(\Sigma, n)$, $S =\frac{1}{n}\bA$ and $I$ be the $p\times p$ identity matrix. \citet{ledoit2004well} look for an optimal linear combination $\Sigma^* = \rho_1 I +\rho_2 S$ to estimate the population covariance matrix; the optimality criterion is described in the following result:
	\begin{thm}{[\citet{ledoit2004well}]}\label{thm:ledoit}
	Let $\|\cdot\|^2$ be the squared Frobenius norm, and let $\langle\cdot,\cdot\rangle$ be its corresponding inner product. Consider the optimization problem:
	\begin{align*}
	\min_{\rho_1,\rho_2} \E(\| \Sigma^* - \Sigma \|^2), \quad \mbox{such that} \quad \Sigma^* = \rho_1 I +\rho_2 S,
	\end{align*}
	where the coefficients $\rho_1,\rho_2$ are nonrandom. Its solution verifies:
	\begin{align*}
	\Sigma^* &= \frac{\beta^2\mu}{\delta^2}I + \frac{\alpha^2}{\delta^2}S,\\
	\E(\| \Sigma^* - \Sigma \|^2) &= \frac{\alpha^2\beta^2}{\delta^2},
	\end{align*}
	where
	$$ \mu = \langle\Sigma, I\rangle,\alpha^2=\|\Sigma -\mu I\|^2, \beta^2 = \E(\| S - \Sigma \|^2),\mbox{ and }\delta^2 = \E(\| S - \mu I \|^2).$$
	\end{thm}

	Furthermore, \citet{ledoit2004well} provide consistent estimators for all quantities $\mu,\alpha^2,\beta^2,\delta^2$ under mild conditions that hold true for normal random variables, and therefore hold true in our setting. The resulting estimator for $\Sigma$ is denoted $S^*$, and we thus get a linearly shrunk $\bA^* = nS^*$. 	
	
\begin{figure}
		\centering
		\includegraphics[width=\linewidth]{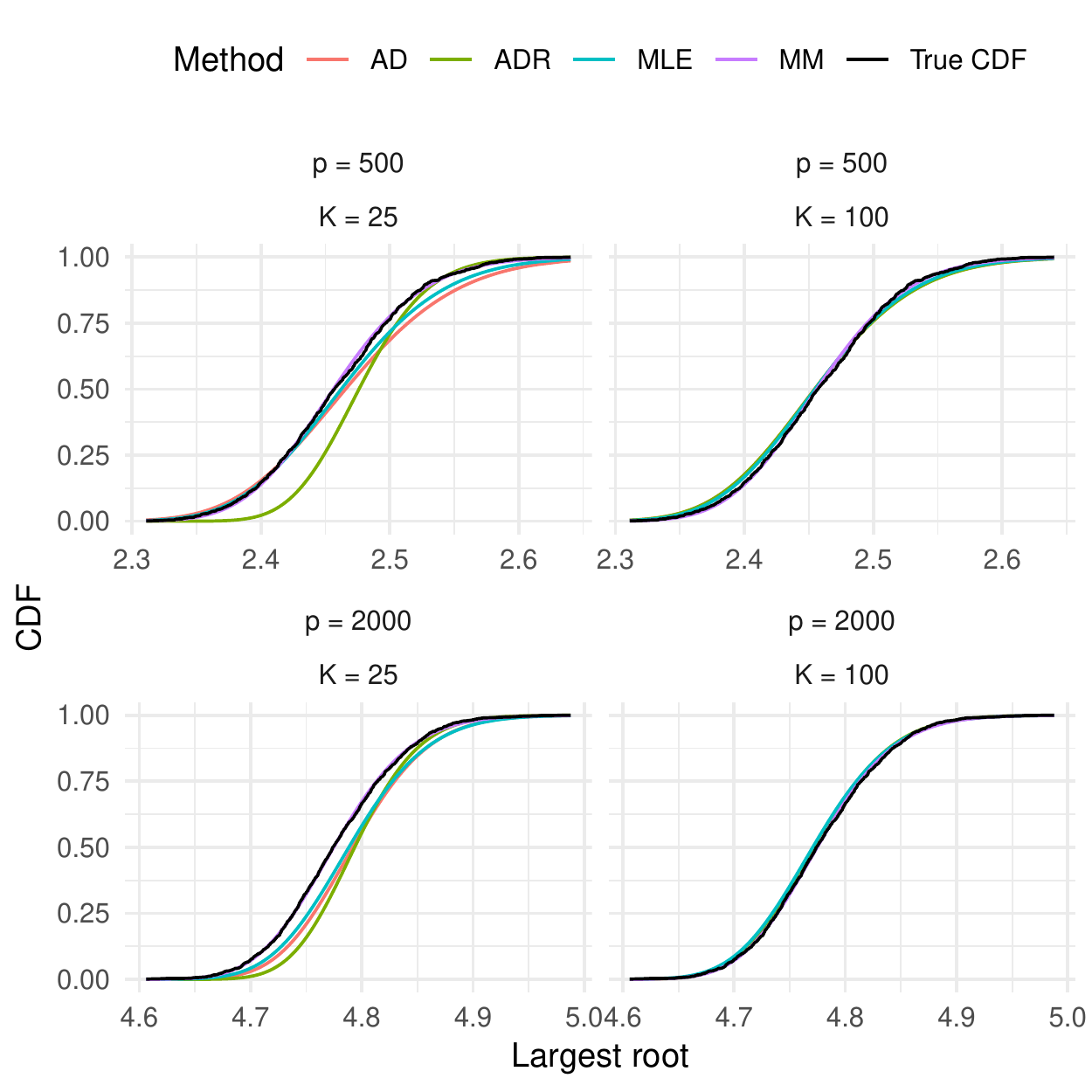}
		\caption{\textbf{Approximation to the CDF under a linearly shrunk $\bA$:} Heuristic, using four different values for the number of permutations, compared to the true CDF}
		\label{fig:approxShrink}
	\end{figure}

	To assess the performance of our Tracy-Widom empirical estimator under this extended setting, we repeated the simulations from Section~\ref{ss:approx} but by substituting the matrix $\bA$ with its linearly shrunk equivalent $\bA^*$. The results appear in Figure~\ref{fig:approxShrink}, and they are very similar to the earlier results; we get good agreement even with small values of $K$, and the method of moments provides a better approximation than the other fitting procedures.
	
	\section{Discussion}
	
	In this work, we investigated the singular double Wishart problem, which arises in the multivariate analysis of high-dimensional datasets. We presented an empirical estimator of the distribution of the largest root that is simple, efficient, and valid for high-dimensional data. Through simulation studies, we showed how our approach leads to a good approximation of the true distribution, and we also showed that it leads to valid p-values. Finally, using a pathway-based approach, we analysed the relationship between DNA methylation and disease type in patients with systemic auto-immune rheumatic diseases. Our analysis used the empirical estimator in two settings: a test for the equality of covariance matrices, and with the dimension-reduction method known as PCEV.
	
	The empirical estimator we presented fills a gap in high-dimensional multivariate analysis. Many common methods, such as MANOVA and CCA, fit into the framework of double Wishart problems. However, classical hypothesis testing breaks down in high dimensions, and therefore analysts often rely on computationally intensive resampling techniques to perform valid inference. For example, genomic studies often require significance thresholds of the order of $10^{-6}$ and lower in order to correct for multiple testing. In this context, a permutation procedure would require at least 1 million resamples. By relying on results from random matrix theory, we can drastically cut down the required number of permutations. Since we only need to estimate two parameters from a location-scale family, good approximation is achieved with less than 100 permutations. Critically, this number is independent of the number of tests performed, and therefore the computation time is reduced by several orders of magnitude.
	
	
	We motivated our empirical estimator of the CDF using an approximation theorem of~\citet{johnstone2008multivariate}. Our approach is further motivated by several results from random matrix theory that suggest a central-limit-type theorem for random matrices, with the Tracy-Widom distribution replacing the normal distribution~\citep{deift2007universal}. More evidence in support of our empirical estimator is given from the study of the largest root of Wishart variates. \emph{On the one hand, \citet{srivastava2007multivariate} and~\citet{srivastava2006multivariate} showed that the asymptotic distribution of $\lambda_{\max}$ is approximately equal to the distribution of the largest eigenvalue of a scaled Wishart matrix.} On the other hand, several results analogous to Johnstone's theorem were also derived for Wishart distributions. Indeed, it has been shown that if $\lambda_1$ is the largest root of a Wishart variate, then there exists $\mu,\sigma$, both functions of the parameters of the Wishart distribution, such that the ratio $\frac{\lambda_1 - \mu}{\sigma}$ converges in distribution to $TW(1)$~\citep{johansson2000shape,johnstone2001distribution,elkaroui2003largest,tracy2009distributions}. 
	
	Finally, the results from Section~\ref{s:ledoit} show that the domain of applicability of our empirical estimator extends beyond that of a strict double Wishart problem. Specifically, we showed that we can replace the matrix $\bA$ in Equation~\ref{eqn:deteqn} by a linearly shrunk estimator based on earlier work by \citet{ledoit2004well}. This approach can easily be applied to multivariate analysis approaches for which $\bA$ is explicitly the covariance matrix of a multivariate random variable. Examples include the test of equality of covariance matrices and PCEV. However, more care is required in order to use these results with CCA.

	
%
%
	
	\bibliographystyle{plainnat}
	\bibliography{reference}
	
	
\end{document}